\documentclass[nohyper, pdftex]{JHEP3}

\usepackage{amsmath,amssymb}
\usepackage{afterpage}
\usepackage{graphicx}
\usepackage{ifthen}
\usepackage{cite}

\graphicspath{{Figs-EPS/}{Figs-PDF/}}

\makeatletter
\DeclareRobustCommand\recite[1]{\begingroup\@fileswfalse\cite{#1}\endgroup}
\makeatother

\newcommand{\Dmq}{\Delta m^2}
\newcommand{\Nuc}[2][]{{\ensuremath{\ifthenelse{\equal{#1}{}}{}{\mbox{}^{#1}}\text{#2}}}}
\newcommand{\eVq}{\ensuremath{\text{eV}^2}}
\newcommand{\PAGEFIGURE}[1]{\FIGURE[!p]{#1}\afterpage\clearpage}

\title{Updated global fit to three neutrino mixing: status of the
  hints of $\theta_{13} > 0$}

\author{M.~C.~Gonzalez-Garcia\\
  C.N.~Yang Institute for Theoretical Physics\\
  State University of New York at Stony Brook\\
  Stony Brook, NY 11794-3840, USA,\\
  {\rm and:}
  Instituci\'o Catalana de Recerca i Estudis Avan\c{c}ats (ICREA),\\
  Departament d'Estructura i Constituents de la Mat\`eria and
  Institut de Ciencies del Cosmos,
  Universitat de Barcelona, Diagonal 647, E-08028 Barcelona, Spain\\
  E-mail:~\email{concha@insti.physics.sunysb.edu}}

\author{Michele Maltoni\\
  Instituto de F\'{\i}sica Te\'orica UAM/CSIC,
  Facultad de Ciencias, Universidad Aut\'onoma de Madrid,
  Cantoblanco, E-28049 Madrid, Spain\\
  E-mail:~\email{michele.maltoni@uam.es}}

\author{Jordi Salvado\\
  Departament d'Estructura i Constituents de la Mat\`eria and
  Institut de Ciencies del Cosmos,
  Universitat de Barcelona, 647 Diagonal, E-08028 Barcelona, Spain\\
  E-mail:~\email{jsalvado@ecm.ub.es}}

\abstract{We present an up-to-date global analysis of solar,
  atmospheric, reactor and accelerator neutrino data in the framework
  of three-neutrino oscillations.  We discuss in detail the
  statistical significance of the observed ``hint'' of non-zero
  $\theta_{13}$ in the solar sector at the light of the latest
  experimental advances, such as the Borexino spectral data, the lower
  value of Gallium rate recently measured in SAGE, and the low energy
  threshold analysis of the combined SNO phase I and phase II. We also
  study the robustness of the results under changes of the inputs such
  as the choice of solar model fluxes and a possible modification of
  the Gallium capture cross-section as proposed by SAGE.  In the
  atmospheric sector we focus on the latest results for $\nu_e$
  appearance from MINOS and on the recent Super-Kamiokande results
  from the combined phases I, II and III, and we discuss their impact
  on the determination of $\theta_{13}$.  Finally, we combine all the
  data into a global analysis and determine the presently allowed
  ranges of masses and mixing.}

\preprint{IFT-UAM/CSIC-10-04\\YITP-SB-10-02}

\keywords{neutrino oscillations, solar and atmospheric neutrinos}

\begin{document}

\section{Introduction}

It is now an established fact that neutrinos are massive and leptonic
flavors are not symmetries of Nature~\cite{Pontecorvo:1967fh,
  Gribov:1968kq}.  In the last decade this picture has become fully
proved thanks to the upcoming of a set of precise experiments. In
particular, the results obtained with solar and atmospheric neutrinos
have been confirmed in experiments using terrestrial beams: neutrinos
produced in nuclear reactors and accelerators facilities have been
detected at distances of the order of hundreds of
kilometers~\cite{GonzalezGarcia:2007ib}.

The minimum joint description of all the neutrino data requires mixing
among all the three known neutrinos ($\nu_e$, $\nu_\mu$, $\nu_\tau$),
which can be expressed as quantum superpositions of three massive
states $\nu_i$ ($i=1,2,3$) with masses $m_i$.  This implies the
presence of a leptonic mixing matrix in the weak charged current
interactions~\cite{Maki:1962mu, Kobayashi:1973fv} which can be
parametrized as:
\begin{equation}
  \label{eq:matrix}
  U =
  \begin{pmatrix}
    c_{12} c_{13}
    & s_{12} c_{13}
    & s_{13} e^{-i\delta_\text{CP}}
    \\
    - s_{12} c_{23} - c_{12} s_{13} s_{23} e^{i\delta_\text{CP}}
    & \hphantom{+} c_{12} c_{23} - s_{12} s_{13} s_{23} e^{i\delta_\text{CP}}
    & c_{13} s_{23} \hspace*{5.5mm}
    \\
    \hphantom{+} s_{12} s_{23} - c_{12} s_{13} c_{23} e^{i\delta_\text{CP}}
    & - c_{12} s_{23} - s_{12} s_{13} c_{23} e^{i\delta_\text{CP}}
    & c_{13} c_{23} \hspace*{5.5mm}
  \end{pmatrix},
\end{equation}
where $c_{ij} \equiv \cos\theta_{ij}$ and $s_{ij} \equiv
\sin\theta_{ij}$.  In addition to the Dirac-type phase
$\delta_\text{CP}$, analogous to that of the quark sector, there are
two physical phases associated to the Majorana character of neutrinos,
which however are not relevant for neutrino
oscillations~\cite{Bilenky:1980cx, Langacker:1986jv} and are therefore
omitted on the present work.  Given the observed hierarchy between the
solar and atmospheric mass-squared splittings there are two possible
non-equivalent orderings for the mass eigenvalues, which are
conventionally chosen as
\begin{align}
  \label{eq:normal}
  \Dmq_{21} &\ll (\Dmq_{32} \simeq \Dmq_{31})
  \text{ with } (\Dmq_{31} > 0) \,;
  \\
  \label{eq:inverted}
  \Dmq_{21} &\ll |\Dmq_{31} \simeq \Dmq_{32}|
  \text{ with } (\Dmq_{31} < 0) \,.
\end{align}
As it is customary we refer to the first option,
Eq.~\eqref{eq:normal}, as the \emph{normal} scheme, and to the second
one, Eq.~\eqref{eq:inverted}, as the \emph{inverted} scheme; in
this form they correspond to the two possible choices of the sign of
$\Dmq_{31}$.  In this convention the angles $\theta_{ij}$ can be taken
without loss of generality to lie in the first quadrant, $\theta_{ij}
\in [0, \pi/2]$, and the CP phase $\delta_\text{CP} \in [0, 2\pi]$.

Within this context, $\Dmq_{21}$, $|\Dmq_{31}|$, $\theta_{12}$, and
$\theta_{23}$ are relatively well
determined~\cite{GonzalezGarcia:2007ib, Fogli:2009zza, Schwetz:2008er,
  Maltoni:2008ka}, while only an upper bound is derived for the mixing
angle $\theta_{13}$ and barely nothing is known on the CP phase
$\delta_\text{CP}$ and on the sign of $\Dmq_{31}$.  Apart from the
importance from a theoretical point of view, establishing whether
$\theta_{13}$ is zero or not is an essential step in the development
of the search strategies for the upcoming experiments.  Once we know
that the two mixing angles $\theta_{12}$ and $\theta_{23}$ are
relatively large, the possibility of experimentally accessing leptonic
CP violation crucially depends on the value of the angle
$\theta_{13}$.  Also a non-zero $\theta_{13}$ is a fundamental
ingredient for a feasible determination of the neutrino mass ordering.
For this reason, it is a main objective of upcoming reactor and
accelerator experiments to directly measure this parameter.  In this
respect, Refs.~\cite{Fogli:2008jx, Fogli:2009ce} pointed out that two
independent hints in favor of a non-zero value of $\theta_{13}$ emerge
from the combination of solar and long-baseline (LBL) reactor data as
well as from the combination of atmospheric, short-baseline reactor
and LBL accelerator data.  Since these signals are the results of
synergies between different data samples, it is particularly important
to verify their stability with respect to new experimental data as
well as to variations on the assumptions in the analysis.

In this work we present the results of an up-to-date global analysis
of solar, atmospheric, reactor and LBL accelerator neutrino data in
the context of three-neutrino oscillations. In Sec.~\ref{sec:solar} we
focus on the solar sector and we assess the stability of the
oscillation parameters with respect to the inclusion of the new
experimental results and theoretical advances which have become public
during the last year. In this context, we find that many of these
changes indeed lower the statistical significance of a non-zero value
of $\theta_{13}$. In Sec.~\ref{sec:atmos} we do the same in the
atmospheric sector, with particular emphasis on the recent $\nu_e$
appearance results from MINOS and the recent Super-Kamiokande results
from the combined phases I, II and III. In Sec.~\ref{sec:global} we
combine all the data together and we determine the presently allowed
ranges of mass and mixing, thus updating our previous results.

\section{Leading $\Dmq_{21}$ oscillations: solar and KamLAND data}
\label{sec:solar}

\FIGURE[!t]{
  \includegraphics[width=0.6\textwidth]{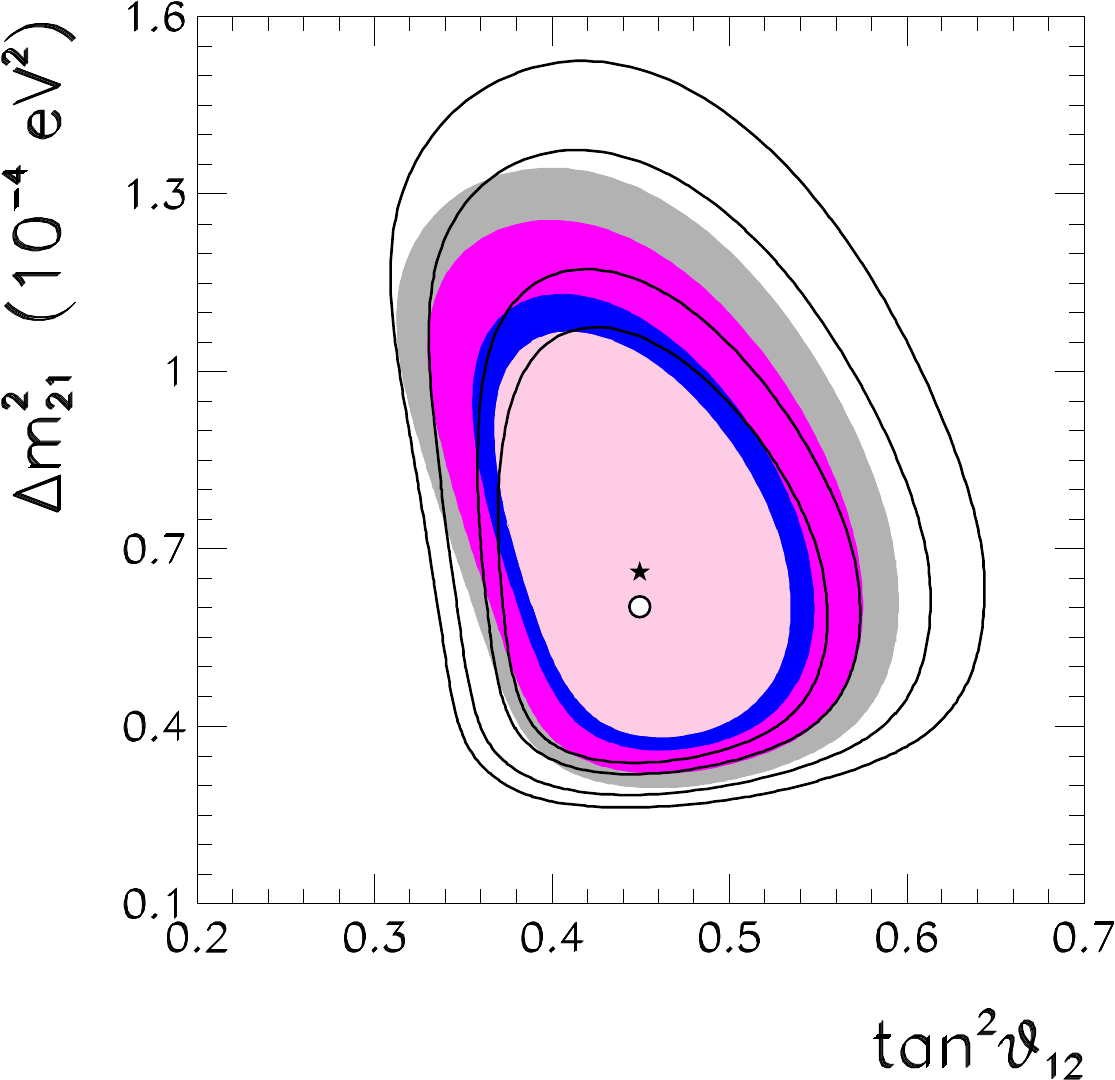}
  \caption{\label{fig:soltwo}
    Allowed parameter regions (at 90\%, 95\%, 99\% and 99.73\% CL for
    2 d.o.f.) from the combined analysis of solar data for
    $\theta_{13}=0$.  The best-fit point is marked with a star.  For
    comparison we also show as empty regions (the best-fit is marked
    by a circle) the results prior to the inclusion of the latest Ga
    capture rate of SAGE~\recite{Abdurashitov:2009tn}, the energy
    spectrum of Borexino~\recite{Arpesella:2008mt,
      Collaboration:2008mr} and the low energy threshold analysis of
    the combined SNO phase I and phase
    II~\recite{Collaboration:2009gd}. In both analysis we use as
    inputs the GS98 solar model fluxes and the Gallium capture
    cross-section of Bahcall~\recite{Bahcall:1997eg}.}}

In the analysis of solar neutrino experiments we include the total
rates from the radiochemical experiments
Chlorine~\cite{Cleveland:1998nv},
Gallex/GNO~\cite{Kaether:2010ag}\footnote{Note that the value
  $R_\text{Gallex} = 73.4 ^{+6.1}_{-6.0}\, \text{(stat)}\,
  ^{+3.7}_{-4.1}\, \text{(syst)}$~SNU presented
  in~\cite{Kaether:2010ag} fully coincide with the preliminary result
  quoted in~\cite{Hahn:2008zz}, and has been commonly used in the
  literature since 2008.}  and SAGE~\cite{Abdurashitov:2009tn}. For
real-time experiments we include the 44 data points of the electron
scattering (ES) Super-Kamiokande phase I (SK-I) energy-zenith
spectrum~\cite{Hosaka:2005um} and the data from the three phases of
SNO~\cite{Aharmim:2007nv, Aharmim:2005gt, Aharmim:2008kc}, including
the results on the low energy threshold analysis of the combined SNO
phase I and phase II~\cite{Collaboration:2009gd} (which we label
SNO-LETA). We also include the main set of the 192 days of Borexino
data~\cite{Arpesella:2008mt} (which we label Borexino-LE) as well as
their high-energy spectrum from 246 live
days~\cite{Collaboration:2008mr} (Borexino-HE). We consider the
following variations on the analysis:
\begin{itemize}
\item{\bf Updated capture rate in gallium.}  In
  Ref.~\cite{Abdurashitov:2009tn} the Russian-American experiment SAGE
  has presented the results of the combined analysis of 168
  extractions (until December~2007). The capture rate of solar
  neutrinos above the 233~keV threshold is
  \begin{equation}
    R^{09}_\text{SAGE} = 65.4^{+3.1}_{-3.0}\, \text{(stat)}\,
    ^{+2.6}_{-2.8}\, \text{(syst)}~\text{SNU}.
  \end{equation}
  which is slightly lower (but fully consistent) than the previous
  result quoted in Ref.~\cite{Hahn:2008zz}, $66.2^{+3.3}_{-3.2}\,
  \text{(stat)}^{+3.5}_{-3.2}\, \text{(syst)}~\text{SNU}$, and
  presents a considerable improvement of the systematic uncertainties.

\item{\bf Possible modification of the capture cross-section in
  gallium.}  Ref.~\cite{Abdurashitov:2009tn} also presents the results
  of a new calibration of the SAGE detector with a reactor-produced
  \Nuc[37]{Ar} neutrino source.  The ratio of observed to expected
  event rate in this experiment, once combined with the measured rates
  in the three prior \Nuc[51]{Cr} neutrino-source experiments with
  Gallium, is $0.87 \pm 0.05$.  As a possible explanation for this low
  result, in Ref.~\cite{Abdurashitov:2009tn} it is proposed that the
  cross-section for neutrino capture by the two lowest-lying
  excited states in \Nuc[71]{Ge} may have been overestimated in
  Ref.~\cite{Bahcall:1997eg}.  As an alternative, the authors consider
  a modified capture cross-section where the contribution from these
  two excited states is set to zero.

\item{\bf Inclusion of the Borexino spectral data.}  Following the
  procedure described in Ref.~\cite{GonzalezGarcia:2009ya} (see
  Appendix A of that work for details) we have included in the
  analysis the 160 data points of the Borexino energy spectrum in the
  $365$--$2000$~keV energy range~\cite{Arpesella:2008mt} as well as
  the 7 points of the high-energy spectrum from the 246 live days of
  Borexino~\cite{Collaboration:2008mr}. In our approach the overall
  normalizations of the \Nuc[11]{C}, \Nuc[14]{C}, \Nuc[210]{Bi} and
  \Nuc[85]{Kr} backgrounds are introduced as free parameters and are
  fitted to the data.

\item{\bf Low energy threshold analysis of the combined SNO phase I
  and phase II data.}  In Ref.~\cite{Collaboration:2009gd} the SNO
  Collaboration reported the results from a joint analysis of their
  Phase~I and Phase~II data with an effective electron kinetic energy
  threshold of $T_\text{eff} = 3.5$~MeV.  Besides the inclusion of the
  lower energy data the analysis present improvements in the
  calibration and analysis techniques to reduce the threshold and
  increase the precision of the results.

  From the point of view of any reanalysis of their data, there is an
  important difference with respect to the previous higher threshold
  results. In Refs.~\cite{Aharmim:2007nv, Aharmim:2005gt} the results
  were quoted in the form of binned (in energy and time) event rates.
  In particular, 34 data points of the day-night spectrum were given
  for SNO-I, and the separate day and night rates for neutral current
  (NC) and ES events as well as the day-night energy-spectrum for
  charge current (CC) events were given for SNO-II.  Instead, in
  Ref.~\cite{Collaboration:2009gd} the collaboration presents its
  reanalysis as a total \Nuc[8]{B} neutrino flux plus an effective
  description (under the assumption of unitarity for active neutrinos)
  of the $\nu_e$ survival probability, whose dependence on $E_\nu$ is
  parametrized as a quadratic function for $P_{ee}^\text{day}$ and a
  linear function for the day/night asymmetry. In other words, the
  SNO-LETA results are given as best-fit, uncertainties and
  correlations of six effective parameters: the \Nuc[8]{B} flux, three
  polynomial coefficients of $E_\nu$ for the $P_{ee}^\text{day}$ and
  two for the day-night asymmetry.  When using these data to perform a
  fit to neutrino oscillations, for each value of the oscillation
  parameters one must first obtain the polynomial coefficients that
  best represent the corresponding
  Mikheev, Smirnov and Wolfenstein (MSW)~\cite{Wolfenstein:1977ue,
    Mikheev:1986gs} oscillation probability.  This has to be done
  taking into account the sensitivity of the SNO detector, and the
  relevant information is provided in the Appendix A of
  Ref.~\cite{Collaboration:2009gd}.  We have followed the procedure
  outlined there and we have verified that we can perfectly reproduce
  their oscillation results in Fig.~37.\footnote{However we notice
    that this procedure does not allow the inclusion of the SNO low
    energy threshold data in analysis in terms of more exotic
    scenarios in which either unitarity in the active neutrino sector
    does not hold (like for scenarios with sterile neutrinos) or the
    energy dependence of the oscillation probability cannot be well
    represented by a simple quadratic function.}

\item{\bf Uncertainties on the Solar Model.}  Recent detailed
  determination of the abundances of the heavy elements on the solar
  surface~\cite{newcomp, Asplund:2009fu} lead to lower values than
  previous studies~\cite{Grevesse:1998bj}.  Any solar model which
  incorporates such lower metallicities fails at explaining the
  helioseismological observations~\cite{Bahcall:2004yr}.  Changes in
  the Sun modeling, in particular of the less known convective zone,
  are not able to account for this discrepancy~\cite{Chaplin:2007uh,
    Basu:2006vh}.  So far there has not been a successful solution of
  this puzzle, so that at present there is no fully consistent
  Standard Solar Model (SSM).  This lead to the construction of two
  different sets of SSM's, one (labeled ``GS'') based on the older
  solar abundances~\cite{Grevesse:1998bj} leading to high metallicity,
  and one (labeled ``AGS'') assuming lower
  metallicity~\cite{Bahcall:2004pz}.  We use the most recent
  recalculation of the fluxes in these two sets from
  Ref.~\cite{Serenelli:2009yc}, and following their notation we refer
  to them as GS98 and AGSS09.
\end{itemize}

\FIGURE[!t]{
  \includegraphics[width=0.8\textwidth]{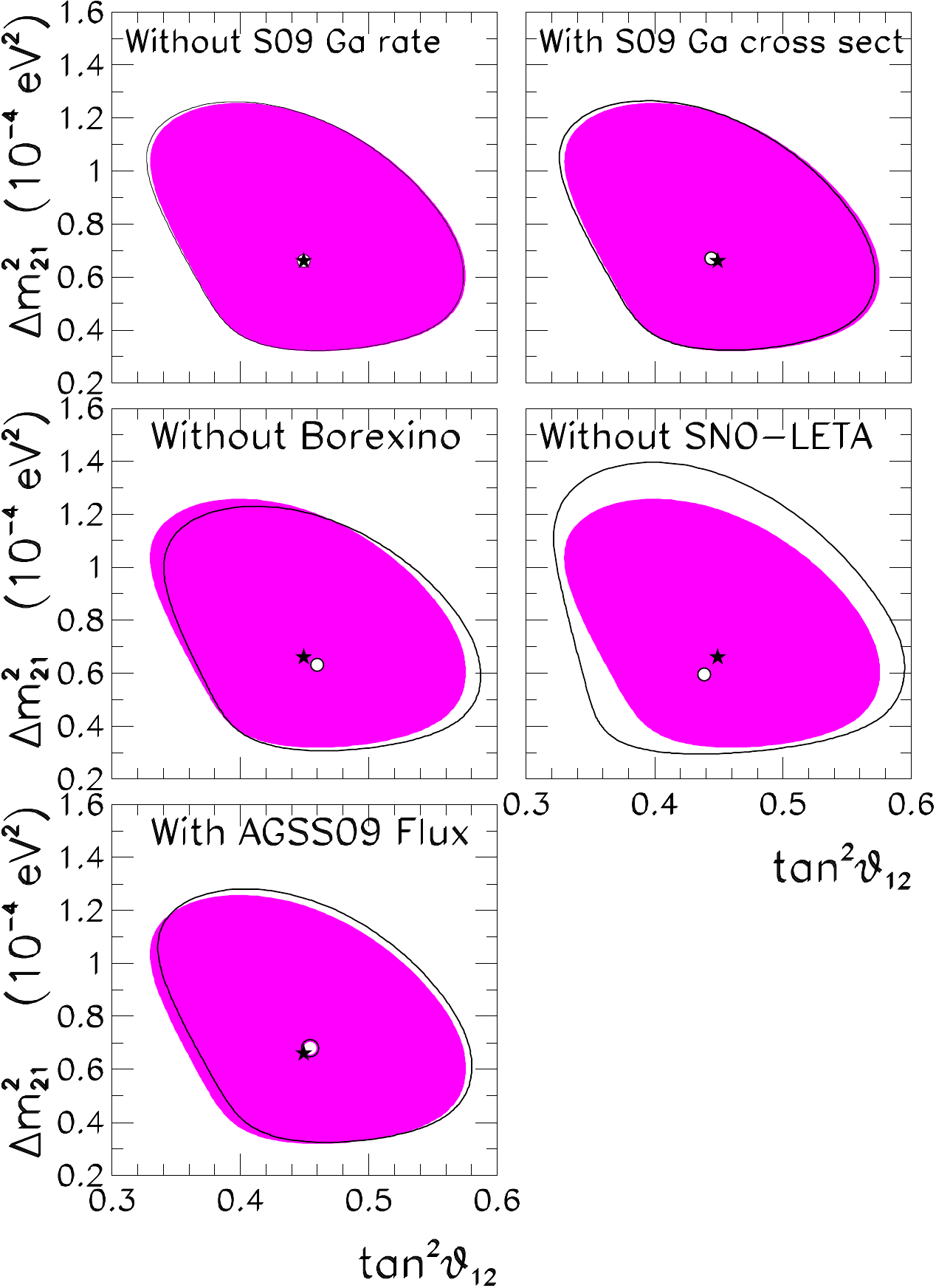}
  \caption{\label{fig:solvary}
    Effects of the different assumed inputs on the 99\% CL parameter
    region from the combined analysis of solar data for
    $\theta_{13}=0$. The full region corresponds to the results show
    in Fig.~\ref{fig:soltwo}. The solid line corresponds to the
    modification due to the change on one of the inputs as labeled in
    each panel.}}

Let us first consider the case $\theta_{13} = 0$.  In
Fig.~\ref{fig:soltwo} we show the present determination of the leading
parameters $\Dmq_{21}$ and $\theta_{12}$ from the updated oscillation
analysis of the solar neutrino data described above in the context of
the GS98 solar model.  For comparison we also show the results
obtained prior to the inclusion of the latest Ga capture rate of
SAGE~\cite{Abdurashitov:2009tn}, the energy spectrum of
Borexino~\cite{Arpesella:2008mt, Collaboration:2008mr} and the
SNO-LETA results~\cite{Collaboration:2009gd} for the same solar model.
As seen in this figure, the inclusion of these results lead to an
improvement on the determination of both $\theta_{12}$ and $\Dmq_{21}$
and for this last one the best-fit value slightly increases.
The effect of the different variations in the analysis is displayed in
Fig.~\ref{fig:solvary}, where we show the 99\% CL region in the
$(\Dmq_{21},\tan^2\theta_{12})$ plane from the present analysis and
the corresponding one when some of the input data or assumption are
modified.  The figure shows that the most quantitatively relevant new
information arises from the inclusion of the SNO-LETA results.  The
inclusion of Borexino tends to shift the region towards slightly lower
values of $\theta_{12}$ angle. Conversely, if the analysis is done in
the context of the AGSS09 model the region is shifted towards slightly
larger $\theta_{12}$.

\subsection{Impact of $\theta_{13}$ on the solar analysis}

\FIGURE[!t]{
  \includegraphics[width=0.8\textwidth]{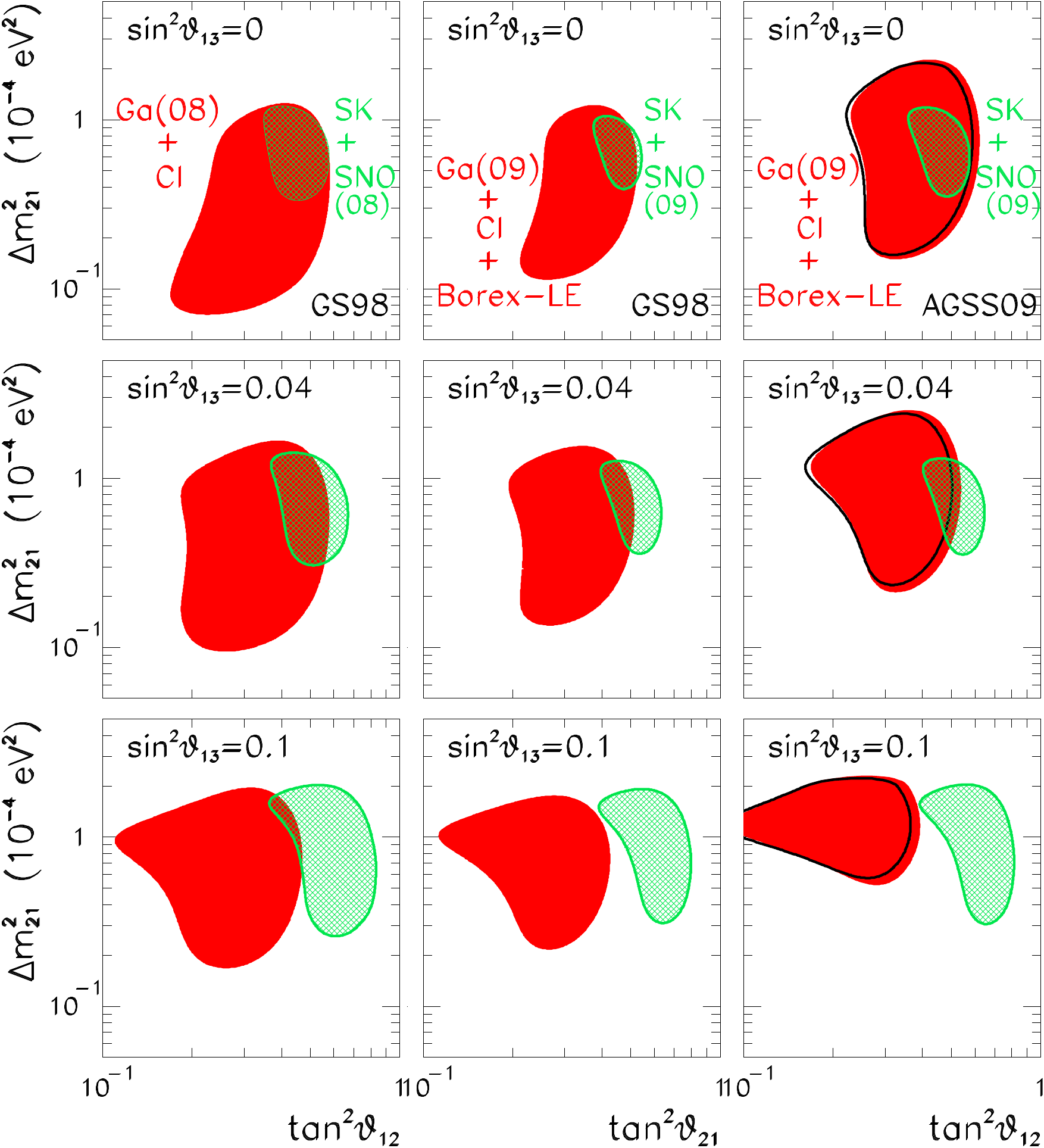}
  \caption{\label{fig:solthree}
    Dependence on $\theta_{13}$ of the allowed ($\Dmq_{21}$,
    $\tan^2\theta_{12}$) regions (at 95\% CL for 2 d.o.f.) from the
    partial analysis of the ``low energy'' and ``high energy'' solar
    neutrino data. The left column corresponds to the analysis prior
    to the inclusion of the latest Ga capture rate of
    SAGE~\recite{Abdurashitov:2009tn}, the energy spectrum of
    Borexino~\recite{Arpesella:2008mt, Collaboration:2008mr} and the
    low energy threshold analysis of the combined SNO phase I and
    phase II~\recite{Collaboration:2009gd}.  In the central and right
    columns the results of those experiments are included. The central
    column corresponds to GS98 solar model fluxes and Gallium capture
    cross-section of Bahcall~\recite{Bahcall:1997eg}.  In the right
    panels the AGSS09 solar model fluxes are used. The full (void)
    regions are obtained with Gallium capture cross-section of
    Bahcall~\recite{Bahcall:1997eg} (modified cross-section in
    Ref.~\recite{Abdurashitov:2009tn}).}}

The survival probability of solar neutrinos in the framework of three
neutrino oscillations can be written as:
\begin{equation}
  \label{eq:ps3}
  P^{3\nu}_{ee} = \sin^4\theta_{13} + \cos^4\theta_{13}
  P^{2\nu}_{ee}(\Dmq_{21},\theta_{12}) \,,
\end{equation}
where we have used the fact that $L^\text{osc}_{31} = 4\pi E /
\Dmq_{31}$ is much shorter than the distance between the Sun and the
Earth, so that the oscillations related to ${L_{31}^\text{osc}}$ are
averaged.  In presence of matter effects
$P^{2\nu}_{ee}(\Dmq_{21},\theta_{12})$ should be calculated taking
into account the evolution in an effective matter density
$n^\text{eff}_{e} = n_e \cos^2 \theta_{13}$.  For $10^{-5}\lesssim
\Dmq/\eVq \lesssim 10^{-4}$, $P^{2\nu}_{ee}(\Dmq_{21},\theta_{12})$
presents the following asymptotic behaviors~\cite{Goswami:2004cn}:
\begin{align}
  \label{eq:ps2l}
  P^{2\nu}_{ee}(\Dmq_{21},\theta_{12})
  &\simeq 1 - \frac{1}{2} \sin^2(2\theta_{12})
  & \text{for}\quad
  E_\nu &\lesssim \text{few} \times 100~\text{KeV}
  \\
  \label{eq:ps2h}
  P^{2\nu}_{ee}(\Dmq_{21},\theta_{12})
  &\simeq \sin^2(\theta_{12})
  & \text{for}\quad
  E_\nu &\gtrsim \text{few} \times 1~\text{MeV}
\end{align}

The impact of the inclusion of a non-zero value of $\theta_{13}$ in
the solar analysis is shown in Fig.~\ref{fig:solthree} and in the
upper-left panel of Fig.~\ref{fig:chisq}.  From this last plot we see
that solar neutrino data by themselves favor $\theta_{13}=0$, although
their sensitivity is very weak for $\sin^2\theta_{13} \lesssim 0.03$.
This behavior can be understood from Fig.~\ref{fig:solthree}, where we
show the allowed regions (at 95\% CL) in the $(\Dmq_{21},
\tan^2\theta_{12})$ plane as obtained from the analysis of low-energy
(radiochemical and Borexino-LE) and high-energy (SK, SNO and
Borexino-HE) solar experiments, for different values of $\theta_{13}$.
As described in Eq.~\eqref{eq:ps3}, for fixed values of $\Dmq_{21}$
and $\theta_{12}$, the inclusion of a small value of $\theta_{13}$
results into a decrease on the predicted rates at a given solar
neutrino experiment. This decrease can be compensated by a shift of
$\Dmq_{21}$ and $\theta_{12}$ which lead to an increase of
$P^{2\nu}_{ee}$.  However the sign of the shift strongly depends on
the characteristic energy of the detected neutrinos. For experiments
detecting neutrinos with energies low enough for matter effects to be
irrelevant (such as Chlorine and Gallium experiments) $P^{2\nu}_{ee}$
is given by Eq.~\eqref{eq:ps2l} and increases as $\theta_{12}$
decreases. Conversely, for experiments detecting neutrinos mostly in
the regime of adiabatic matter oscillations (such as SK and SNO)
$P^{2\nu}_{ee}$ is given by Eq.~\eqref{eq:ps2h} and increases as
$\theta_{12}$ increases. Consequently the combined fit worsens with
$\theta_{13}$.

\FIGURE[!t]{
  \includegraphics[width=0.95\textwidth]{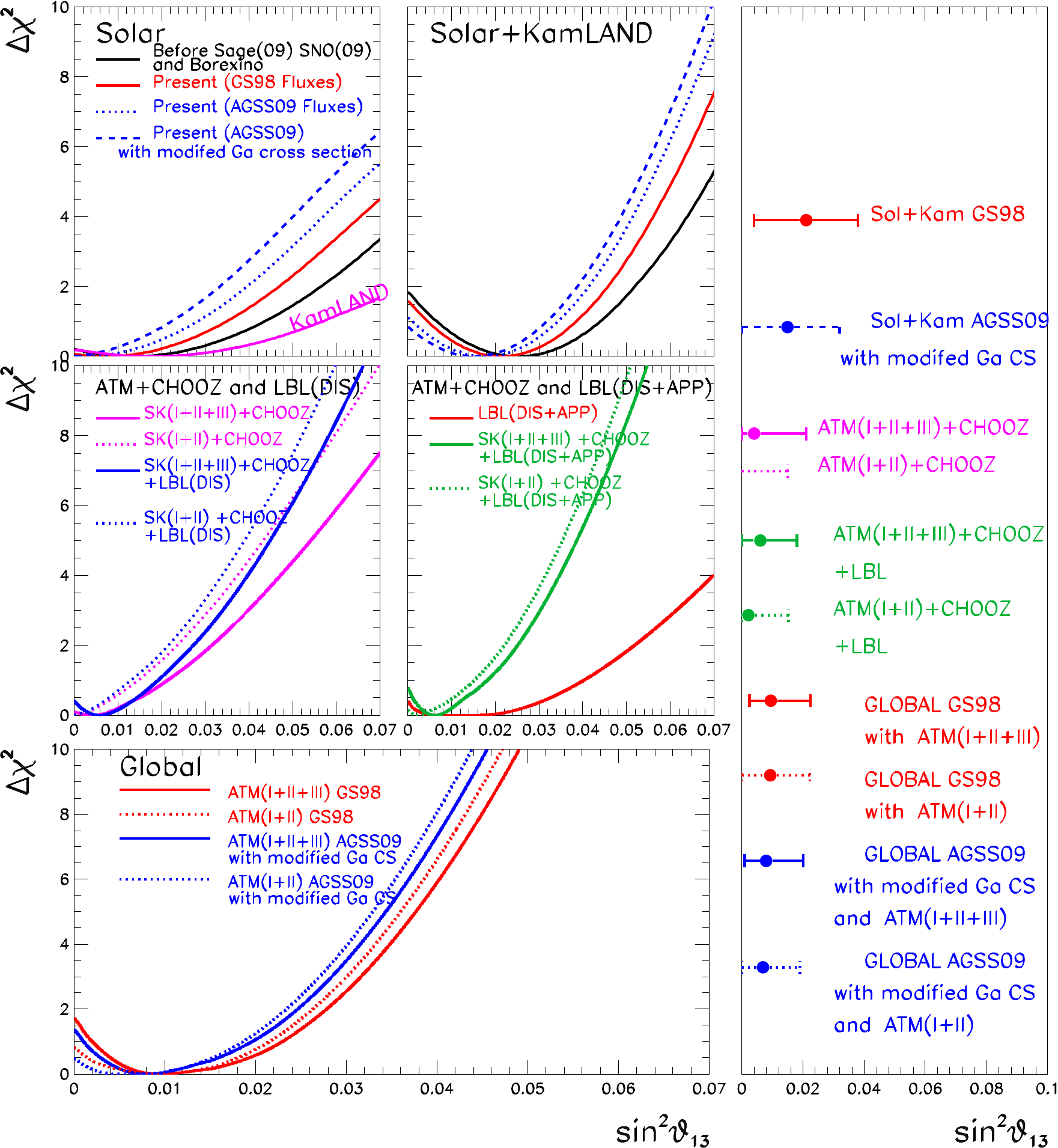}
  \caption{\label{fig:chisq}
    Dependence of $\Delta\chi^2$ on $\sin^2\theta_{13}$ for the
    different analysis as labeled in the figure and the corresponding
    $1\sigma$ ranges.}}

Fig.~\ref{fig:solthree} also illustrates how the inclusion of the new
data strengthens this tension. The quantitative improvement on the
corresponding bound on $\theta_{13}$ is displayed in the upper-left
panel of Fig.~\ref{fig:chisq}.  We have also studied the dependence of
this result on the assumed solar model and on the possible
modification of the neutrino capture in gallium.  We find that the
AGSS09 fluxes lead to a stronger bound on $\theta_{13}$ and the same
happens with the modified gallium cross-section. This second effect
can be easily understood as follows: because of the lower
cross-section, a slightly larger survival probability is required to
fit the same data. This shifts the $(\Dmq_{21},\tan^2\theta_{12})$
allowed region obtained from the analysis of the radiochemical solar
experiments towards slightly lower values of $\theta_{12}$ (as
explicitly shown in the last column of Fig.~\ref{fig:solthree}). As a
consequence it increases the tension with the SNO+SK favored mixing
angle for non-zero $\theta_{13}$.

\subsection{Combination with KamLAND and the hint of $\theta_{13} \neq 0$}

\FIGURE[!t]{
  \includegraphics[width=0.8\textwidth]{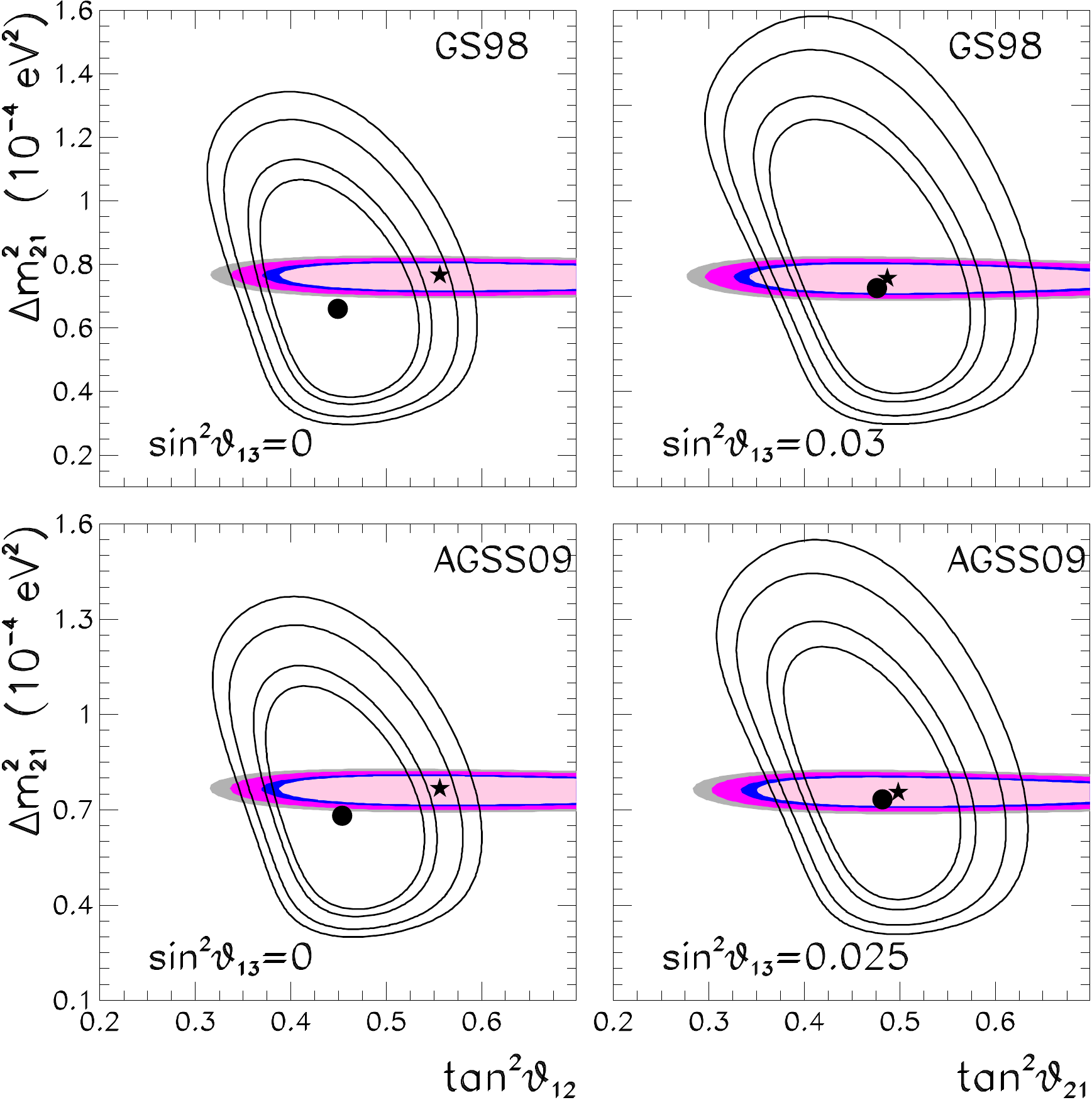}
  \caption{\label{fig:solkam}
    Allowed parameter regions (at 90\%, 95\%, 99\% and 99.73\% CL for
    2 d.o.f.) from analysis of KamLAND (full regions with best-fit
    marked by a star) and solar (void regions with best-fit marked by
    a dot) data for two values of $\theta_{13}$ as labeled in the
    figure and for the two solar models.}}

We show in the left panels of Fig.~\ref{fig:solkam} the present
determination of the leading parameters $\Dmq_{21}$ and $\theta_{12}$
(for $\theta_{13}=0$) from the analysis of KamLAND spectral
data~\cite{Shimizu:2008zz} compared to those from the updated solar
analysis for the two solar models considered.  While the results show
perfect agreement in $\Dmq_{21}$, there appears to be a mismatch in
the favored value of $\theta_{12}$ as determined from KamLAND compared
to the one from solar neutrinos, this last one being mostly sensitive
to the precise value of CC/NC event (\textit{i.e.}, to $\langle
P_{ee}\rangle \propto \sin^2\theta_{12})$ as determined by SK and SNO.

As it was pointed out in Ref.~\cite{Fogli:2008jx} and widely discussed
in the literature~\cite{Fogli:2009zza, Schwetz:2008er,
  Maltoni:2008ka,Balantekin:2008zm},
this mismatch can be lifted by a non-zero value of
$\theta_{13}$.  This happens because, as discussed above, the CC/NC
event rate can be fitted with a higher value of $\theta_{12}$ provided
that a non-zero $\theta_{13}$ is included. Conversely for KamLAND
Eq.~\eqref{eq:ps3} also holds with
\begin{equation}
  P_{ee}^{2\nu, \text{kam}} = 1 - \frac{1}{2}\sin^2(2\theta_{12})
  \sin^2\frac{\Dmq_{21} L}{2E} \,.
\end{equation}
So for $\theta_{13}>0$ the KamLAND spectrum can be well fitted with a
smaller value of $\theta_{12}$, and consequently the best-fit values
for solar and KamLAND analysis agree better for $\theta_{13}\neq
0$. This behavior is clearly visible in the right panels of
Fig.~\ref{fig:solkam} and in the upper-central panel of
Fig.~\ref{fig:chisq}. As mentioned before, the best-fit value of
$\theta_{12}$ for solar neutrino fit within the AGSS09 model is
slightly larger than for the GS98 model, and therefore the required
value of $\theta_{13}$ to achieve agreement with KamLAND is smaller
for the AGSS09 model.

However, one must notice that the better agreement between the solar
and KamLAND analysis for $\theta_{13}\neq 0$ has to be contrasted with
the worsening of the global description of the solar neutrino data
previously described.  As a consequence of this tension we find that
the inclusion of the new data and of the modified gallium capture
cross-section also tend to lower the best-fit value of $\theta_{13}$
as well as its corresponding statistical significance (see
upper-central panel of Fig.~\ref{fig:chisq}).  Altogether we find that
the $1\sigma$ range for $\theta_{13}$ as determined from the global
analysis of solar and KamLAND data changes from the old value
$\sin^2\theta_{13}=0.025\pm 0.018$ (before the inclusion of SNO-LETA,
SAGE-09 and Borexino data) to:
\begin{equation}
  \sin^2\theta_{13} =
  \begin{cases}
    0.021 \pm 0.017 &\text{for GS98,}
    \\
    0.017 \pm 0.017 &\text{for AGSS09,}
    \\
    0.015 \pm 0.017 &\text{for AGSS09 with modified \Nuc{Ga}
      cross-section.}
  \end{cases}
\end{equation}

\section{Leading $\Dmq_{31}$ oscillations: atmospheric, CHOOZ and
  accelerator data}
\label{sec:atmos}

\FIGURE[!t]{
  \includegraphics[width=0.95\textwidth]{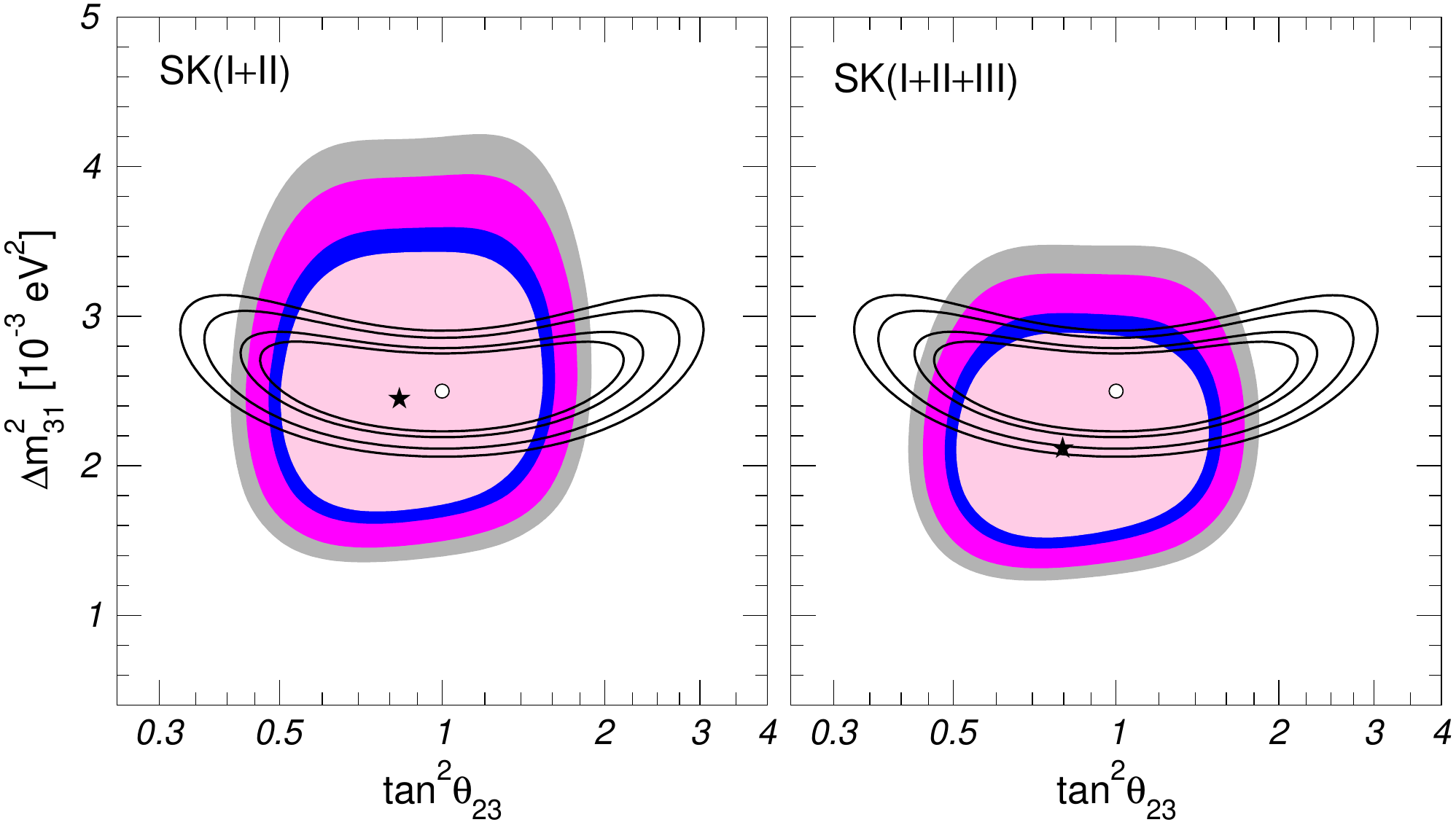}
  \caption{\label{fig:atmtwo}
    Allowed parameter regions (at 90\%, 95\%, 99\% and 99.73\% CL for
    2 d.o.f.) from the analysis of atmospheric data (full regions,
    best-fit marked with a star) and LBL data (void regions, best-fit
    marked by a circle) for $\theta_{13} = 0$ and $\Dmq_{21} = 7.6
    \times 10^{-5}~\eVq$.}}

In this section we present two different analyses of the atmospheric
data. The first one is very similar to the one detailed in
Ref.~\cite{GonzalezGarcia:2007ib}, and includes the results from the
first run of Super-Kamiokande, which accumulated data from May 1996 to
July 2001 (1489 day exposure) and is usually referred as
SK-I~\cite{Ashie:2005ik}, as well as the data obtained with the
partial coverage after the 2001 accident (804 day exposure), the
so-called SK-II period~\cite{atmskII}. We will refer to this analysis
as SK(I+II).
The second one is based on the new analysis recently presented by the
Super-Kamiokande collaboration including also the data taken from
December 2005 to June 2007, usually referred as
SK-III~\cite{Wendell:2010md}.  Apart from the inclusion of these new
event rates, in this data release the selection criteria and the
corresponding estimate of uncertainties for the SK-I and SK-II periods
have been changed with respect to the previous SK(I+II) analysis. We
have therefore performed a reanalysis of the new combined samples from
phases I, II and III as presented in~\cite{Wendell:2010md}. We refer
to the results of this analysis as SK(I+II+III).
It is important to point out that already since SK-II the
Super-Kamiokande collaboration has been presenting its experimental
results in terms of a large number of data samples.  The rates for
some of those samples cannot be theoretically predicted (and therefore
include in a statistical analysis) without a detailed simulation of
the detector which can only be made by the experimental collaboration
itself. Thus our results represent the most up-to-date analysis of
the atmospheric neutrino data which can be performed outside the
collaboration.  For details on our simulation of the data samples and
the statistical analysis see the Appendix of
Ref.~\cite{GonzalezGarcia:2007ib}.

For what concerns LBL accelerator experiments, we combine the results
on $\nu_\mu$ disappearance from K2K~\cite{Ahn:2006zza} with those
obtained by MINOS at a baseline of 735~km after a two-year exposure to
the Fermilab NuMI beam, corresponding to a total of $3.36 \times
10^{20}$ protons on target~\cite{Adamson:2008zt}.  We also include the
recent results on $\nu_\mu\rightarrow \nu_e$ transitions based on an
exposure of $7 \times 10^{20}$ protons on target~\cite{minapp70fnal,
  minapp70slac}.

In order to test the description of the present data in the absence of
$\theta_{13}$-induced effects we show in Fig.~\ref{fig:atmtwo} the
present determination of the leading parameters $\Dmq_{31}$ and
$\theta_{23}$ for $\theta_{13}=0$ and $\Dmq_{21}=7.6 \times
10^{-5}~\eVq$ from the two atmospheric neutrino analyses and the LBL
accelerator results. For concreteness we plot only normal ordering;
the case of inverted ordering gives practically identical results as
long as $\theta_{13} = 0$. This figure illustrates how the bounds on
the oscillation parameters $\theta_{23}$ and $\Dmq_{31}$ emerges from
a complementarity of atmospheric and accelerator neutrino data:
$|\Dmq_{31}|$ is determined by the spectral data from MINOS, whereas
the mixing angle $\theta_{23}$ is still largely dominated by
atmospheric data from Super-Kamiokande with a best-fit point close to
maximal mixing.
It is important to note that there is a very good agreement in the
location of the best-fit points from SK(I+II) and MINOS. This is not
the case for SK(I+II+III) for which the best-fit point in
$|\Dmq_{31}|$ is now lower than the one obtained from LBL. In what
follows we will discuss the impact of this tension on the allowed
range of $\theta_{13}$.

\subsection{Impact of $\theta_{13}$ on the atmospheric and
  LBL $\nu_\mu$ disappearance data}

\FIGURE[!t]{
  \includegraphics[width=0.95\textwidth]{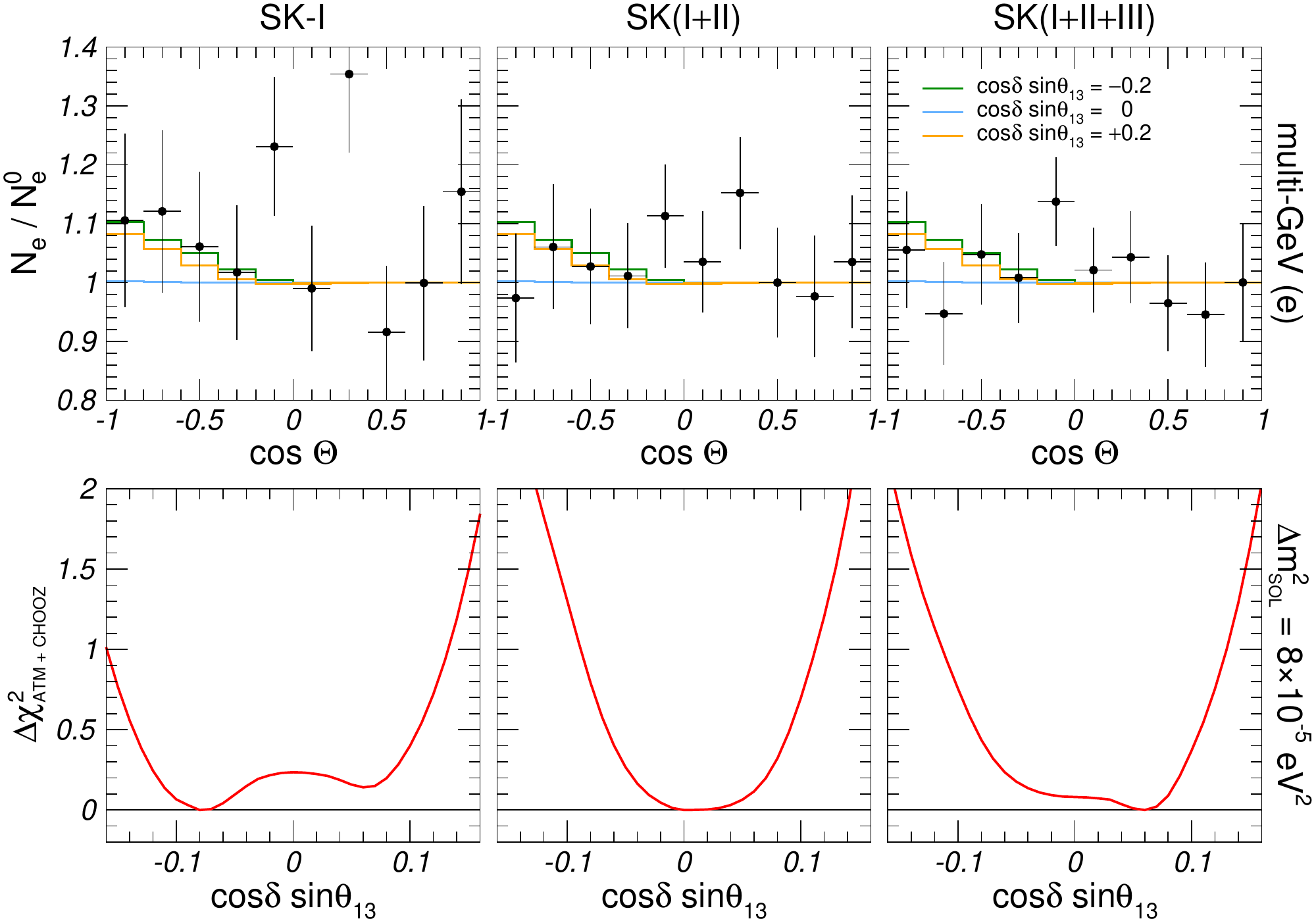}
  \caption{\label{fig:atmrate}
    Zenith distribution for multi-GeV $e$-like events (upper panels),
    and $\Delta\chi^2$ dependence on $\cos\delta_\text{CP}
    \sin\theta_{13}$ marginalized over all undisplayed parameters
    (lower panels).  The fits include the CHOOZ data as well as the
    full atmospheric data samples for SK-I (left panels), SK-I with
    modified multi-GeV $e$-like data (central-left panels), SK(I+II)
    (central-right panels) and SK(I+II+III) (right panels).}}

It is well known that a very important contribution to our knowledge
of $\theta_{13}$ arises from the negative results on $\bar\nu_e$
disappearance at short baselines at the CHOOZ reactor
experiment~\cite{Apollonio:1999ae}. Given the values of $\Dmq_{21}$
obtained from solar and KamLAND experiments, the disappearance of
$\bar\nu_e$ at the CHOOZ distances arises from oscillations due to
$|\Dmq_{31}|$ whose amplitude is proportional to
$\sin^2(2\theta_{13})$.  Consequently, when combined with the bounds
on $|\Dmq_{31}|$ from atmospheric and $\nu_\mu$ disappearance at LBL
experiments the CHOOZ result implies an upper bound on $\theta_{13}$.

In Ref.~\cite{Fogli:2008jx} a hint for a non-zero value of
$\theta_{13}$ was obtained at $0.9\sigma$ from the analysis of
atmospheric data from SK-I, $\nu_\mu$ disappearance at long-baseline
experiments and CHOOZ.  The authors traced its origin to subleading
effects driven by $\Dmq_{21}$.  This result was extensively studied in
Ref.~\cite{Maltoni:2008ka}, where it was pointed out that the
appearance of the hint is triggered by an excess of multi-GeV $e$-like
data in the first two angular bins of SK-I data, which lead to a
better fit for this sample in the case of a non-zero value of
$\theta_{13}$.  This conclusion is summarized in
Fig.~\ref{fig:atmrate}: as clearly visible in the leftmost panel,
there is a small but definite preference for a non-zero value of
$\theta_{13}$ in SK-I. Once the SK-II data are included in the
analysis the event rates of those bins are considerably reduced, and
as a consequence no hint of non-zero $\theta_{13}$ is found in
SK(I+II).  As can be seen in the rightmost panels, in the new
SK(I+II+III) data release there is again a small excess in the first
angular bin of of multi-GeV $e$-like data, and consequently a weak
preference for non-zero $\theta_{13}$ reappears. Such ``hint'',
however, is considerably weaker than in the SK-I case and completely
negligible from the statistical point of view.

In order to better understand the impact of the various data samples,
in the second row of Fig.~\ref{fig:chisq} we plot the bound on
$\theta_{13}$ as implied by different combinations of atmospheric and
LBL experiments. In all panels the results are marginalized with
respect to the undisplayed parameters, and in particular to the sign
of $\Dmq_{31}$.  As can be seen, for both SK(I+II) and SK(I+II+III)
the combined analysis of atmospheric and CHOOZ show no relevant
preference for nonzero $\theta_{13}$, and the data results exclusively
into a bound on $\theta_{13}$.  Such bound is weaker for SK(I+II+III)
as a result of the lower values of $|\Dmq_{31}|$ preferred by the
atmospheric analysis, and gets slightly strengthened by the inclusion
of LBL $\nu_\mu$ disappearance data.  Since the $\nu_\mu$ survival
probability at MINOS and K2K is practically insensitive to such small
values of $\theta_{13}$, this improvement arises as an indirect effect
driven by the better determination of $|\Dmq_{31}|$.
On the other hand, it should be noted that when combining the new
SK(I+II+III) with LBL $\nu_\mu$ disappearance data a small preference
for a non-zero value of $\theta_{13}$ arises. This effect is a
consequence of the mismatch between the best-fit $|\Dmq_{31}|$ in
atmospheric and MINOS data shown in Fig.~\ref{fig:atmtwo}. Within our
atmospheric neutrino analysis this is only a $\Delta\chi^2 = 0.5$
effect. However, the analysis of the atmospheric data presented in
Ref.~\cite{Wendell:2010md} by Super-Kamiokande seems to exhibit a
non-negligible dependence of the preferred $\Dmq_{31}$ range on
$\theta_{13}$ (see, \textit{e.g.}, the leftmost panels of Figs.~7 and
8 in~\cite{Wendell:2010md}).  It would therefore be very interesting
if the Super-Kamiokande and MINOS collaborations could perform a
combined analysis of their samples in order to fully establish the
statistical significance of this effect.

\subsection{$\nu_\mu \rightarrow \nu_e$ appearance results in MINOS}

\FIGURE[!t]{
  \includegraphics[width=0.95\textwidth]{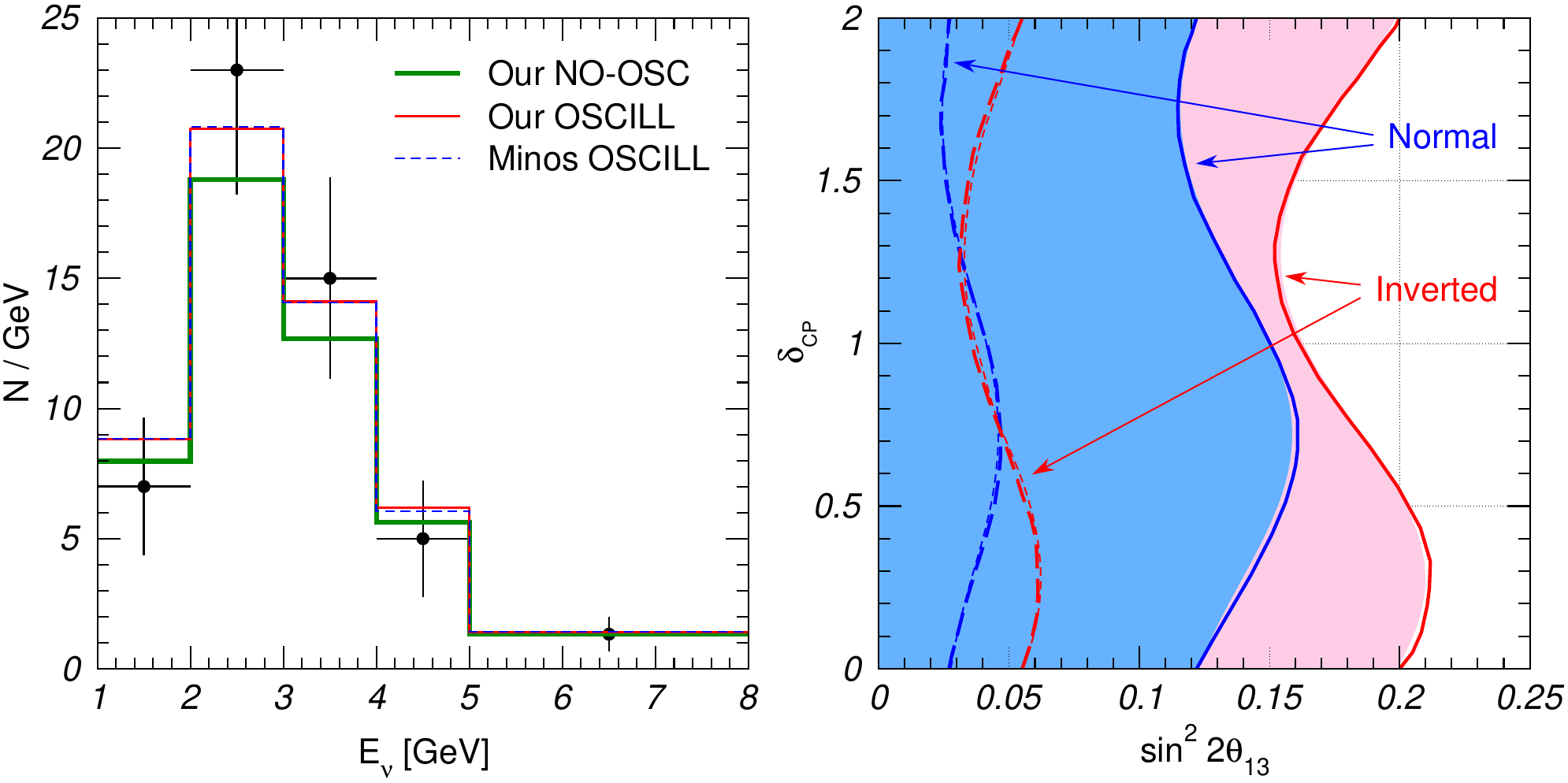}
  \caption{\label{fig:minapp}
    Left: reconstructed energy distribution of the $\nu_e$ charged
    current events in the MINOS far detector.  We show the background
    prediction (thick solid green line) as well as our (red solid) and
    MINOS (blue dashed) predictions for the expected spectrum in the
    presence of $\nu_\mu\rightarrow \nu_e$ oscillations with
    $\sin^22\theta_{13}=0.115$ and $\delta_\text{CP} = 0$. Right:
    best-fit and upper bound on $\theta_{13}$ as a function of
    $\delta_\text{CP}$ for the MINOS $\nu_e$ appearance data. The
    dashed (dotted) lines represent the best-fit as obtained from our
    (MINOS) analysis for normal and inverted ordering. The solid
    regions are our 90\% bounds for normal (darker blue) and inverted
    (lighter red) ordering; the solid lines show the corresponding
    limits from MINOS, taken from Ref.~\recite{minapp70fnal}. In both
    panels the remaining parameters are fixed to
    $\tan^2\theta_{12}=0.45$, maximal $\theta_{23}$, $\Dmq_{21}=7.6
    \times 10^{-5}~\eVq$ and $|\Dmq_{32}|=2.43 \times 10^{-3}~\eVq$.}}

In Ref.~\cite{Collaboration:2009yc} the MINOS collaboration reported
their first results on the search for $\nu_\mu\rightarrow\nu_e$
transitions based on an exposure of $3.14 \times 10^{20}$
protons-on-target in the Fermilab NuMI beam.  They observed 35 events
in the Far Detector with a background of $27\pm 5\, \text{(stat)} \pm
2\, \text{(syst)}$ events predicted by their measurements in the Near
Detector.  This result corresponded to a $1.5\sigma$ excess which
could be explained by a non-zero value of $\theta_{13}$.  Recently a
new analysis with double statistics (exposure of $7 \times 10^{20}$)
has been presented~\cite{minapp70fnal, minapp70slac}.  The MINOS
collaboration reported the observation of 54 events with an expected
background of $49.1\pm 7.0\, \text{(stat)} \pm 2.7\, \text{(syst)}$,
thus reducing the excess above background to $0.7\sigma$.

In the left panel of Fig.~\ref{fig:minapp} we show the reconstructed
energy distribution of the observed events together with the
background expectations.  We compare our simulation of the expected
spectrum in the presence of $\nu_\mu \rightarrow \nu_e$ oscillations
with the one given by the MINOS collaboration. In the right panel of
Fig.~\ref{fig:minapp} we perform an oscillation analysis of the MINOS
total event rate, and again we compare our limits on $\theta_{13}$ for
given $\delta_\text{CP}$ to those obtained by the MINOS
collaboration. In both cases we find perfect agreement.

In the central panel of Fig.~\ref{fig:chisq} we show the information
on $\theta_{13}$ from the analysis of the LBL data from both $\nu_e$
appearance and $\nu_\mu$ disappearance and its impact when combined
with the atmospheric and CHOOZ data.
As seen in this figure, the inclusion of MINOS appearance data in
combination with either atmospheric neutrino samples leads to a
stronger upper bound on $\theta_{13}$. Due to the $0.7\sigma$ excess
it also increases slightly the significance for $\theta_{13}\neq 0$.
As expected, this increase is larger when combined with SK(I+II+III),
since in this case the MINOS $\nu_e$ appearance excess adds to the
$\Dmq_{31}$ mismatch effect between SK(I+II+III) and MINOS $\nu_\mu$
disappearance discussed above.
Altogether we find that the $1\sigma$ range for $\theta_{13}$ as
determined from the analysis of atmospheric, CHOOZ and LBL data reads:
\begin{align}
  \sin^2\theta_{13} &= 0.002^{+0.013}_{-0.005}
  && \text{CHOOZ} + \text{LBL(DIS+APP)} + \text{SK(I+II)} \,,
  \\
  \sin^2\theta_{13} &= 0.006^{+0.012}_{-0.007}
  && \text{CHOOZ} + \text{LBL(DIS+APP)} + \text{SK(I+II+III)} \,,
\end{align}
where for the sake of illustration we have extrapolated the lower
$1\sigma$ bound to the unphysical region $\sin^2\theta_{13} < 0$.

\section{Global results and conclusions}
\label{sec:global}

\PAGEFIGURE{
  \includegraphics[width=0.9\textwidth]{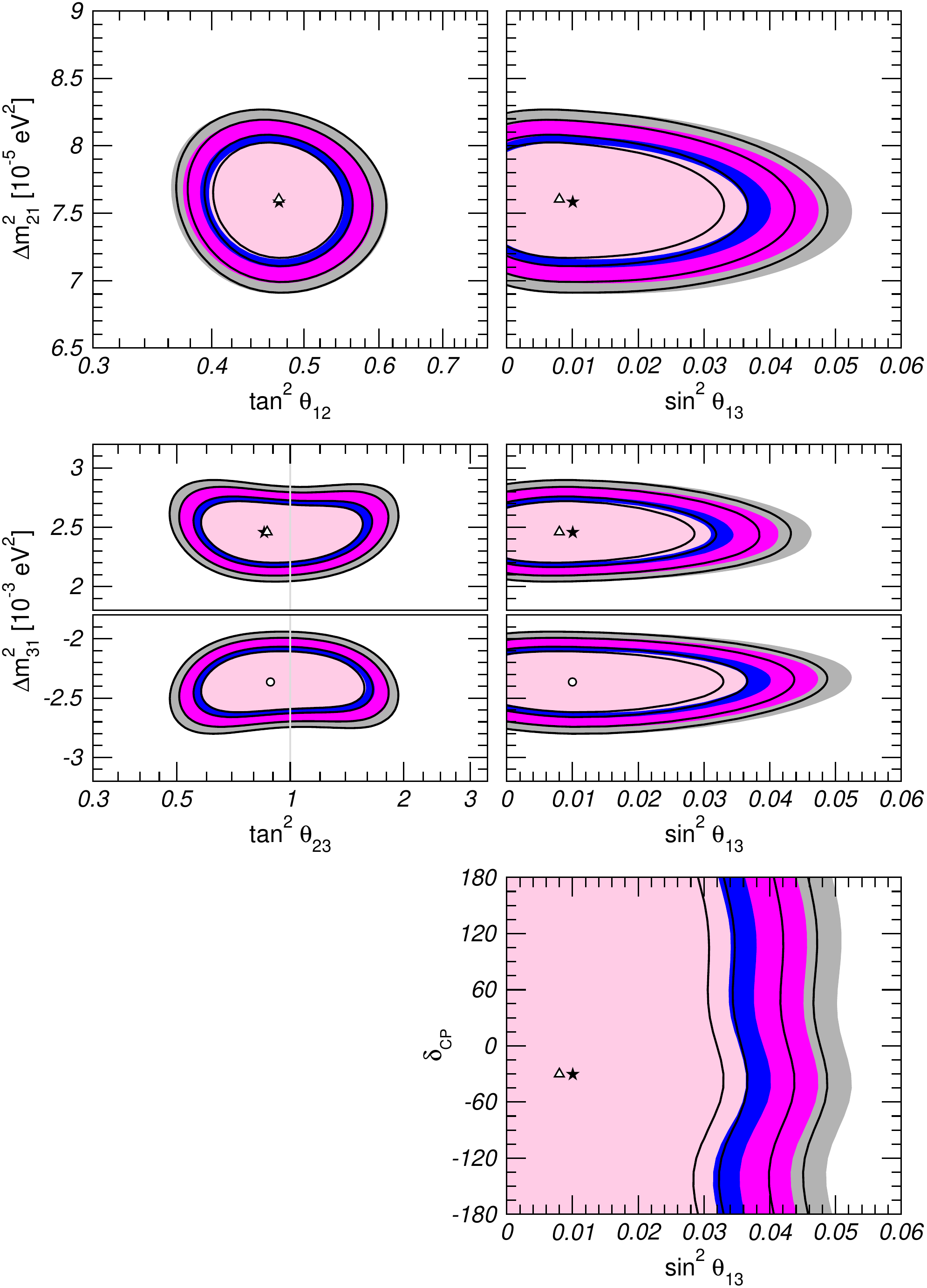}
  \caption{\label{fig:global}
    Global $3\nu$ oscillation analysis. Each panels shows
    two-dimensional projection of the allowed five-dimensional region
    after marginalization with respect to the undisplayed parameters.
    The different contours correspond to the two-dimensional allowed
    regions at 90\%, 95\%, 99\% and $3\sigma$ CL.  The full regions
    correspond to the analysis done in the framework of the GS98 solar
    model and with Ga capture cross-section in
    Ref.~\recite{Bahcall:1997eg} while the void regions correspond the
    analysis with AGSS09 solar fluxes and the modified Ga capture
    cross-section in Ref.~\recite{Abdurashitov:2009tn}.}}

The results of the global combined analysis including the SK(I+II+III)
atmospheric neutrino data and all dominant and subdominant oscillation
effects are summarized in Fig.~\ref{fig:global}, where we show the
different projections of the allowed six-dimensional parameter space.
The full regions correspond to the analysis done in the framework of
the GS98 solar model and with Ga capture cross-section in
Ref.~\cite{Bahcall:1997eg} while the void regions correspond the
analysis with AGSS09 solar fluxes and the modified Ga capture
cross-section in Ref.~\cite{Abdurashitov:2009tn}.  The regions in each
panel are obtained after marginalization of $\chi^2_\text{global}$
with respect to the undisplayed parameters.  In the lower panel we
show the allowed regions in the ($\sin^2\theta_{13}$,
$\delta_\text{CP}$) plane. As seen in the figure, at present the
sensitivity to the CP phase is marginal but we find that the bound on
$\sin^2\theta_{13}$ can vary by about $\sim$ 30\% depending on the
exact value of $\delta_\text{CP}$. This arises mainly from the
interference of $\theta_{13}$ and $\Dmq_{21}$ effects in the
atmospheric neutrino observables, as well as from the new MINOS
$\nu_e$ appearance data.  The derived ranges for the six parameters at
the $1\sigma$ ($3\sigma$) level are:
\begin{equation*}
  \begin{aligned}
    & \text{GS98 with Gallium}
    && \text{AGSS09 with modified}
    \\[-1mm]
    &\text{cross-section from~\cite{Bahcall:1997eg}}
    &&\text{Gallium cross-section~\cite{Abdurashitov:2009tn}}
    \\[+1mm]
    \hline
    \\[-4mm]
    \Dmq_{21}
    &= 7.59 \,\pm 0.20 \,\left(_{-0.69}^{+0.61}\right) \times 10^{-5}~\eVq
    && \text{Same}
    \\
    \Dmq_{31} &=
    \left\lbrace
    \begin{matrix}
      -2.36 \pm 0.11 \,\left(\pm 0.37\right)
      \times 10^{-3}~\eVq  \,
      \\[1mm]
      +2.46 \pm 0.12 \,\left(\pm{0.37}\right)
      \times 10^{-3}~\eVq \,
    \end{matrix}
    \right.
    && \text{Same}
    \\
    \theta_{12}
    &= 34.4       \pm 1.0 \,\left(_{-2.9}^{+3.2}\right)^\circ
    && 34.5 \pm 1.0 \,\left(_{-2.8}^{+3.2}\right)^\circ
    \\
    \theta_{23}
    &= 42.8 \,_{-2.9}^{+4.7} \,\left(_{\hphantom{0}-7.3}^{+10.7}\right)^\circ
    && \text{Same}
    \\
    \theta_{13}
    &= 5.6 \,^{+3.0}_{-2.7} \,\left(\leq 12.5\right)^\circ
    && 5.1 \,^{+3.0}_{-3.3} \,\left(\leq 12.0\right)^\circ
    \\
    \big[ \sin^2\theta_{13}
      &= 0.0095 \,^{+0.013}_{-0.007} \, \left(\leq 0.047 \right) \big]
    &\big[ & 0.008 \,^{+0.012}_{-0.007} \,\left(\leq 0.043\right) \big]
    \\
    \delta_\text{CP} &\in [0,\, 360]
    && \text{Same}
  \end{aligned}
\end{equation*}
For each parameter the ranges are obtained after marginalizing with
respect to to the other parameters.  For $\Dmq_{31}$ the allowed
ranges are formed by two disconnected intervals which correspond to the
two possible mass orderings.  The absolute best-fit lays in the
positive $\Dmq_{31} = +2.46 \times 10^{-3}~\eVq$. The $1\sigma$ and
$3\sigma$ ranges are defined with respect to this absolute minimum.
In particular the local best-fit in the inverse mass ordering,
$\Dmq_{31} = -2.36 \times 10^{-3}~\eVq$, is at a $\Delta\chi^2 =
0.12$.

In summary we have presented the results of an up-to-date global
analysis of neutrino oscillation data including the detailed analysis
of the Borexino spectra, the lower value of Ga rate as lastly measured
in SAGE, the low energy threshold analysis of the combined SNO phase I
and phase II, the results for $\nu_e$ appearance from MINOS and the
recent reanalysis of SK(I+II+III) atmospheric data. We have studied
the robustness of the hints of a non-vanishing value of $\theta_{13}$
under the inclusion of the new data and variations of the assumptions
about the solar model and the possible modification of the
cross-section for neutrino capture in Ga as advocated by the SAGE
collaboration in order to explain their new calibration data.

We found that the inclusion of the new solar data, and in particular
of the SNO-LETA results tends to lower the statistical significance of
$\theta_{13}\neq 0$ while the results from $\nu_e$ appearance from
MINOS and the new SK(I+II+III) atmospheric neutrino reanalysis
increases it.  We conclude that the significance of $\theta_{13}\neq
0$ is:
\begin{equation*}
  \begin{array}{l@{\hspace{7mm}}l@{\hspace{7mm}}l}
    & \text{GS98 with Gallium}
    & \text{AGSS09 with modified}
    \\
    &\text{cross-section from~\cite{Bahcall:1997eg}}
    &\text{Gallium cross-section~\cite{Abdurashitov:2009tn}}
    \\[+1mm]
    \hline
    \\[-4mm]
    \text{solar+KamLAND}
    & \text{CL} = 79\% \; (1.26\sigma)
    & \text{CL} = 70\% \; (1.05\sigma)
    \\[+2mm]
    \parbox{46mm}{%
      $\mbox{} + \text{SK(I+I+III)} + \text{CHOOZ}$\\[-1mm]
      $\mbox{} + \text{LBL(Dis+App)}$}
    & \text{CL} = 81\% \; (1.31\sigma)
    & \text{CL} = 76\% \; (1.17\sigma)
  \end{array}
\end{equation*}
The determination of the other oscillation parameters is rather
robust.

\section*{Acknowledgments}

This work is supported by Spanish MICINN grants 2007-66665-C02-01,
FPA-2009-08958 and FPA-2009-09017 and consolider-ingenio 2010 grant
CSD-2008-0037, by CSIC grant 200950I111, by CUR Generalitat de
Catalunya grant 2009SGR502, by Comunidad Autonoma de Madrid through
the HEPHACOS project S2009/ESP-1473, by USA-NSF grant PHY-0653342 and
by EU grant EURONU.

\bibliographystyle{JHEP}
\bibliography{references}

\end{document}